\documentclass[preprint,aps,showpacs,floats,nofootinbib]{revtex4}
\usepackage{graphicx}
\usepackage{amsmath}
\usepackage{amssymb}
\begin{document}
\title{Radiative and Semileptonic $B$ Decays Involving the Tensor Meson $K_2^*(1430)$
in the Standard Model and Beyond}
\author{Hisaki Hatanaka}
\author{Kwei-Chou Yang}
\affiliation{Department of Physics, Chung-Yuan Christian University, Chung-Li,
Taiwan 320, R.O.C.}
\date{\today}
\begin{abstract}
We study semileptonic and radiative $B$ decays involving the strange tensor
meson $K_2^*(1430)$ in the final state. Using the large energy effective theory
(LEET) techniques, we formulate the $B \to K_2^*$ transition form factors in
large recoil region. All the form factors can be parametrized in terms of two
independent LEET functions $\zeta_\perp$ and $\zeta_\parallel$. The magnitude
of $\zeta_\perp$ is estimated from the data for ${\cal B}(B \to
K_2^*(1430)\gamma)$. Assuming a dipole $q^2$-dependence for the LEET functions
and $\zeta_\parallel/\zeta_\perp = 1.0 \pm 0.2$, for which the former consists
with the QCD counting rules and the latter is favored by the $B\to \phi K_2^*$
data, we investigate the decays $B \to K_2^* \ell^+ \ell^-$ and $B \to K_2^*
\nu \bar{\nu}$, where the contributions due to $\zeta_\parallel$ are suppressed
by $m_{K_2^*}/m_B$. For the $B \to K_2^* \ell^+ \ell^-$ decay, in the large
recoil region where the hadronic uncertainties are considerably reduced, the
longitudinal distribution $d F_L/ d s$ is reduced by $20-30$\% due to the
flipped sign of  $c_7^{\rm eff}$ compared with the standard model result.
Moreover, the forward-backward asymmetry zero is about 3.4 GeV$^2$ in the
standard model, but changing the sign of $c_7^{\rm eff}$ yields a positive
asymmetry for all values of the invariant mass of the lepton pair. We calculate
the branching fraction for $B \to K_2^* \nu \bar{\nu}$ in the standard model.
Our result exhibits the impressed resemblance between $B \to K_2^*(1430)
\ell^+\ell^-$, $\nu \bar{\nu}$ and $B \to K^*(892) \ell^+ \ell^-$,
$\nu\bar{\nu}$.
\end{abstract}
\maketitle
%
\newcommand{ \slashchar }[1]{\setbox0=\hbox{$#1$}   
   \dimen0=\wd0                                     
   \setbox1=\hbox{/} \dimen1=\wd1                   
   \ifdim\dimen0>\dimen1                            
      \rlap{\hbox to \dimen0{\hfil/\hfil}}          
      #1                                            
   \else                                            
      \rlap{\hbox to \dimen1{\hfil$#1$\hfil}}       
      /                                             
   \fi}                                             %
%
%
%
\newcommand{\Tr}{{\mathop{\mbox{Tr}}\nolimits}} 
\newcommand{\tr}{{\mathop{\mbox{tr}}\nolimits}} 
\newcommand{\Det}{{\mathop{\mbox{Det}}\nolimits}} 
\newcommand{\diag}{{\mathop{\mbox{diag}}\nolimits}} 
\newcommand{\Diag}{{\mathop{\mbox{Diag}}\nolimits}} 
\newcommand{\Li}{{\mathop{\mbox{Li}}\nolimits}} 
\renewcommand{\Re}{\mathop{\mbox{Re}}} 
\renewcommand{\Im}{\mathop{\mbox{Im}}} 
\newcommand{\del}{\partial}
\newcommand{\Hc}{\mbox{H.c.}}
\newcommand{\bvec}[1]{{\boldsymbol{#1}}}
%
\newcommand{\calA}{{\cal A}}
\newcommand{\calB}{{\cal B}}
\newcommand{\calC}{{\cal C}}
\newcommand{\calD}{{\cal D}}
\newcommand{\calE}{{\cal E}}
\newcommand{\calF}{{\cal F}}
\newcommand{\calG}{{\cal G}}
\newcommand{\calH}{{\cal H}}
\newcommand{\calI}{{\cal I}}
\newcommand{\calJ}{{\cal J}}
\newcommand{\calK}{{\cal K}}
\newcommand{\calL}{{\cal L}}\renewcommand{\L}{{\cal L}}
\newcommand{\calM}{{\cal M}}
\newcommand{\calN}{{\cal N}}
\newcommand{\calO}{{\cal O}}
\newcommand{\calP}{{\cal P}}
\newcommand{\calQ}{{\cal Q}}
\newcommand{\calR}{{\cal R}}
\newcommand{\calS}{{\cal S}}
\newcommand{\calT}{{\cal T}}
\newcommand{\calU}{{\cal U}}
\newcommand{\calV}{{\cal V}}
\newcommand{\calW}{{\cal W}}
\newcommand{\calX}{{\cal X}}
\newcommand{\calY}{{\cal Y}}
\newcommand{\calZ}{{\cal Z}}
\newcommand{\bra}[1]{{\langle {#1}}}
\newcommand{\ket}[1]{{ {#1} \rangle}}
\newcommand{\bbra}[1]{{\langle {#1} |}}
\newcommand{\bket}[1]{{| {#1} \rangle}}
\newcommand{\SM}{{\rm SM}}
\newcommand{\psibar}{\bar{\psi}}
\newcommand{\barpsi}{\bar{\psi}}
\newcommand{\qbar}{\bar{q}}
\newcommand{\ubar}{\bar{u}}
\newcommand{\dbar}{\bar{d}}
\newcommand{\cbar}{\bar{c}}
\newcommand{\sbar}{\bar{s}}
\newcommand{\tbar}{\bar{t}}
\newcommand{\bbar}{\bar{b}}
\newcommand{\nubar}{\bar{\nu}}
\renewcommand{\l}{\ell}
\newcommand{\lbar}{\bar{\l}}
\newcommand{\eV}{{\rm eV}}
\newcommand{\MeV}{{\rm MeV}}
\newcommand{\GeV}{{\rm GeV}}
\newcommand{\TeV}{{\rm TeV}}
\newcommand{\degree}{^\circ}
\newcommand{\rad}{{\rm rad.}}
\newcommand{\BABAR}{BABAR}
\newcommand{\BELLE}{Belle}
\newcommand{\eps}{\epsilon}
\newcommand{\veps}{\varepsilon}
\newcommand{\mnrs}{\mu\nu\rho\sigma}
\newcommand{\mnr}{\mu\nu\rho}
\newcommand{\munu}{\mu\nu}
\newcommand{\alphabeta}{\alpha\beta}
\newcommand{\toLEET}{\stackrel{LEET}{\longrightarrow}}
\newcommand{\tilden}{\tilde{n}}
\newcommand{\scv}{\slashchar{v}}
\newcommand{\scn}{\slashchar{n}}
\newcommand{\para}{\parallel}
\newcommand{\zetav}{\zeta^{(v)}}
\newcommand{\zetaa}{\zeta^{(a)}}
\newcommand{\zetat}{\zeta^{(t)}}
\newcommand{\zetatf}{\zeta^{(t_5)}}
\newcommand{\hatV}{\hat{V}}
\newcommand{\hatA}{\hat{A}}
\newcommand{\hatT}{\hat{T}}
\newcommand{\hatTf}{{\hat{T}_5}}
\newcommand{\hatS}{\hat{S}}
\newcommand{\hatP}{\hat{P}}
\newcommand{\cdotv}{\cdot v}
\newcommand{\Br}{{\cal B}}
\newcommand{\eff}{{\rm eff}}
\newcommand{\ktwo}{K_2^*}
\newcommand{\ktwov}{\ktwo(1430)}
\newcommand{\ktwop}{K_2^{*+}}
\newcommand{\ktwom}{K_2^{*-}}
\newcommand{\ktwoz}{K_2^{*0}}
\newcommand{\ktwopv}{\ktwop(1430)}
\newcommand{\ktwomv}{\ktwom(1430)}
\newcommand{\ktwozv}{\ktwoz(1430)}
\newcommand{\barktwo}{\overline{K}_2^*}
\newcommand{\barB}{\overline{B}}
\newcommand{\mB}{m_{B}}
\newcommand{\mktwo}{m_{K_2^*}}
\renewcommand{\l}{\ell}
\newcommand{\lp}{\l^+}
\newcommand{\lm}{\l^-}
\newcommand{\lpm}{\l^+\l^-}
\newcommand{\mupm}{\mu^+\mu^-}
\newcommand{\kstar}{K^*}
\newcommand{\kstarp}{K^{*+}}
\newcommand{\kstarv}{K^*(892)}
\newcommand{\kstarpv}{K^{*+}(892)}

\newcommand{\tveps}{\tilde{\veps}}
\newcommand{\alphaem}{\alpha_{EM}}
\newcommand{\alphal}{\alpha_L}
\newcommand{\betat}{\beta_T}
\newcommand{\tzeta}{\tilde{\zeta}}
\newcommand{\absp}{|\vec{p}_{\ktwo}|}
\newcommand{\hc}{\mbox{H.c.}}
\newcommand{\sect}{Sec.~}
\newcommand{\fig}{Fig.~}
\newcommand{\tbl}{Table~}
%
\section{Introduction}\label{Introduction}

\begin{table}[tbp]
\caption{Branching fractions
of radiative and semileptonic $B$ decays involving
$\kstar$ or $\ktwo$ .}\label{ktwo-exp}
\begin{ruledtabular}
\begin{tabular}{llll}
Mode & $\Br$ [$10^{-6}$] &
Mode & $\Br$ [$10^{-6}$]
\\
\hline
$B^+\to K^{*+}(892)\gamma$ & $45.7\pm1.9$ \cite{:2008cy,Nakao:2004th,Coan:1999kh} &
$B^0\to K^{*0}(892)\gamma$ & $44.0\pm1.5$ \cite{:2008cy,Nakao:2004th,Coan:1999kh}
\\
$B^+\to K_2^{*+}(1430)\gamma$ & $14.5\pm4.3$ \cite{Aubert:2003zs} &
$B^0\to K_2^{*0}(1430)\gamma$ & $12.4\pm2.4$ \cite{Aubert:2003zs,Nishida:2002me}
\\
$B^+\to K^{*+}(892)e^+ e^-$ & $1.42^{+0.43}_{-0.39}$ \cite{Aubert:2006vb,Adachi:2008sk}&
$B^0\to K^{*0}(892)e^+ e^-$ & $1.13^{+0.21}_{-0.18}$ \cite{Aubert:2006vb,Adachi:2008sk}
\\
$B^+\to K^{*+}(892)\mu^+ \mu^-$ & $1.12^{+0.32}_{-0.27}$
\cite{Aubert:2006vb,Adachi:2008sk,Anderson:2001nt}& $B^0\to K^{*0}(892)\mu^+
\mu^-$ & $1.00^{+0.15}_{-0.13}$
\cite{Aubert:2006vb,Adachi:2008sk,Anderson:2001nt}
\\
$B^+\to K^{*+}(892)\nu \nubar$ & $ < 80$ \cite{:2007zk,:2008fr}&
$B^0\to K^{*0}(892)\nu \nubar$ & $ < 120$ \cite{:2007zk,:2008fr}
\\
\end{tabular}
\end{ruledtabular}
\end{table}

The flavor-changing neutral current (FCNC) processes involving $b \to s(d)$
transitions occur only at loop-level in the standard model (SM) and thus
provide an important testing ground to look for new physics phenomena.
Radiative $B$ decays can offer bounds on the CKM matrix elements $|V_{ts}|$ and
$|V_{td}|$ as well as powerful constraints on new physics. The absolute value
of $c_7^{\eff}$, which is the Wilson coefficient of electromagnetic dipole
operator, extracted from the current $B \to X_s \gamma$ data is consistent with
the SM prediction within errors.

The $b \to s \lpm$ processes arise from photonic penguin, $Z$-penguin and
$W$-box diagrams. The inclusive $B \to X_s \lpm$ and exclusive $B \to K^{(*)}
\lpm$ decays have been measured \cite{:2008sk,Aubert:2008ps}. We summarize the
current data for branching fractions of exclusive radiative and semileptonic
$B$ decays relevant to the FCNC $b \to s$ transition in \tbl\ref{ktwo-exp}
\cite{Ishikawa:2006fh,Barberio:2008fa,:2008cy,Nakao:2004th,Coan:1999kh,Aubert:2003zs,Nishida:2002me,Aubert:2006vb,Adachi:2008sk,Anderson:2001nt,:2007zk,:2008fr,:2008ju}.
The FCNC processes may receive sizable new-physics contributions
\cite{Burdman:1995ks,Hewett:1996ct,Li:2004vh,Xu:2006vk,Colangelo:2006vm,Chen:2002zk}.
Recently, \BABAR{} and \BELLE{} have measured interesting observables, $\kstar$
longitudinal fraction, forward-backward asymmetry and isospin asymmetry, in the
$B \to \kstar \lpm$ decays
\cite{Ishikawa:2006fh,:2008ju,:2008sk,Aubert:2008ps,Aubert:2006vb,Adachi:2008sk}.
Although the data are consistent with the SM predictions, all measurements
favor the flipped-sign $c_7^\eff$ models \cite{Eigen:2008nz}.  The minimal
flavor violation supersymmetry models with large $\tan\beta$ can be fine-tuned
to have the flipped sign of $c_7^\eff$ \cite{Ali:1999mm,Feldmann:2002iw}, for
which the charged Higgs is dominant. However, the contributions of the charged
Higgs exchange to $c_9$ and $c_{10}$ are suppressed by $1/\tan^2\beta$ for
large $\tan\beta$.

The measurements of inclusive and various exclusive decays relevant to FCNC
transitions can shed light on new physics. We have studied $B \to
K_1(1270)\gamma$ and $B \to K_1(1270) \lpm$
\cite{Hatanaka:2008xj,Hatanaka:2008gu}, where the $K_1(1270)$ is the $P$-wave
meson. $B \to K_1(1270)\gamma$ has been measured by \BELLE \cite{Yang:2004as}.
In this paper, we focus on the exclusive processes $B \to \ktwov \gamma$, $B\to
\ktwov \lpm$ and $B \to \ktwov \nu \nubar$, where $\ktwov$ is the strange
tensor meson with positive parity.

The $B \to \ktwov \gamma$ decays have been observed by \BELLE{} and \BABAR{}
collaborations \cite{Aubert:2003zs,Nishida:2002me}. See also
\tbl\ref{ktwo-exp}. Corresponding semileptonic decays can be expected to see
soon. Because both $\ktwo$ and $\kstar$  mainly decay to the two-body $K \pi$
mode, therefore the angular-distribution analysis for the $B \to K^* \lpm$
decays are applicable to the study for $B \to \ktwo \lpm$ decays.

In experiments, the exclusive mode is much more easier to accessible than the
inclusive process. However, the former contains form factors parametrizing
hadronic matrix elements, and thus suffers from large theoretical
uncertainties. $B \to \ktwo$ transition form factors, which are relevant to the
study of the radiative and semileptonic $B$ decays into a $\ktwo$, are less
understood compared with  $B \to \kstar$ ones. So far only some quark model
results about them \cite{HC,Ebert:2001pc,Datta:2007yk}. In this paper we
formulate the $B \to \ktwo$ form factors in the large recoil region using the
large energy effective theory (LEET) techniques \cite{Charles:1998dr}. We will
show that all the form factors can be parameterized in terms of two independent
form factors $\zeta_\perp$ and $\zeta_\para$ in the LEET limit. The former form
factor can be estimated by using the data for $B \to \ktwov \gamma$, while the
latter only gives corrections of order $m_{\ktwo}/m_B$ in the amplitude.

We study the longitudinal distribution $d F_L/d s$ and forward-backward
asymmetry for the $B \to \ktwo \lpm$ decay. Particularly, we find that in the
large recoil region, where the uncertainties of these observables arising from
the form factors are considerably reduced not only due to taking the ratio of
form factors but also due to the evaluation in the large $E_{\ktwo}$ limit. For
the new-physics effect, we will focus on the possible correction due to the
$c_7^\eff$ with the sign flipped.

We calculate the branching fraction for $B \to \ktwo \nu \nubar$ in the SM.
This mode enhanced by the summation over three light neutrinos is theoretically
cleaner due to the absence of long-distance corrections related to the relevant
four-fermion operators. This decay is relevant for the nonstandard $Z^0$
coupling \cite{Buchalla:2000sk}, light dark matter \cite{Bird:2004ts} and
unparticles \cite{Georgi:2007ek,Aliev:2007gr}.

The paper is organized as follows. In \sect\ref{sec-LEET} we formulate the $B
\to T$ form factors using the LEET techniques. In \sect\ref{NumericalStudy}, we
numerically study the radiative and semileptonic $B$ meson decays into the
$\ktwov$. We conclude with a summary in \sect\ref{Summary}.
%
\section{$B \to T$ form factors in the LEET}\label{BtoT_LEET}\label{sec-LEET}
\newcommand{\epsv}[1]{\veps(#1)}
For simplicity we work in the rest frame of the $B$ meson (with mass $m_B$) and
assume that the light tensor meson $T$ (with mass $m_T$) moves along the
$z$-axis. The momenta of the $B$ and $T$ are given by
\begin{eqnarray}
p_B^\mu = (m_B ,0,0,0) \equiv m_B \, v^\mu,
\quad
p_T^\mu = (E,0,0,p_3) \equiv E \, n^\mu,
\end{eqnarray}
respectively. Here the tensor meson's energy $E$ is given by
\begin{eqnarray}
E = \frac{m_B}{2}\left(1 - \frac{q^2}{m_B^2} + \frac{m_T^2}{m_B^2}\right),
\end{eqnarray}
where $q \equiv p_B - p_T$. In the LEET limit,
\begin{eqnarray}
E, m_B \gg m_T, \Lambda_{{\rm QCD}}, \label{LEETlimit}
\end{eqnarray}
we simply have
\begin{eqnarray}
v^\mu = (1,0,0,0),
\quad
n^\mu \simeq (1,0,0,1).
\end{eqnarray}
The polarization tensors $\veps(\lambda)^{\mu\nu}$ of the massive spin-2 tensor
meson with helicity $\lambda$ can be constructed in terms of the polarization
vectors of a massive vector state \cite{Berger:2000wt}
\begin{eqnarray}
\epsv{0}^{*\mu} = (p_3,0,0,E)/m_T,
\quad
\epsv{\pm}^{*\mu} = (0,\mp1,+i,0)/\sqrt{2},
\end{eqnarray}
and are given by
\begin{eqnarray}
\veps^{\mu\nu}(\pm2) &\equiv& \epsv{\pm}^\mu \epsv{\pm}^\nu,
\\
\veps^{\mu\nu}(\pm1) &\equiv& \sqrt{\frac{1}{2}}
(\epsv{\pm}^\mu \epsv{0}^\nu + \epsv{0}^\mu \epsv{\pm}^\nu),
\\
\veps^{\mu\nu}(0) &\equiv& \sqrt{\frac{1}{6}}
 (\epsv{+}^\mu \epsv{-}^\nu + \epsv{-}^\mu \epsv{+}^\nu)
 + \sqrt{\frac{2}{3}}  \epsv{0}^\mu \epsv{0}^\nu.
\end{eqnarray}

Due to the purpose of the present study, we calculate the $\barB \to T$
transition form factors:
\begin{eqnarray}
\bra{T}|V^\mu|\ket{\barB},\quad
\bra{T}|A^\mu|\ket{\barB},\quad
\bra{T}|T^{\mu\nu}|\ket{\barB},\quad
\bra{T}|T_5^{\mu\nu}|\ket{\barB},
\end{eqnarray}
where $V^\mu \equiv \psibar\gamma^\mu b$, $A^\mu \equiv
\psibar\gamma^\mu\gamma_5 b$, $T^{\mu\nu} = \psibar \sigma^{\mu\nu} b$ and
$T_5^{\mu\nu} = \psibar \sigma^{\mu\nu}\gamma_5 b$. There is a trick to write
down the form factors in the LEET limit. We first note that we have three
independent classes of Lorentz structures (i) $\eps^{\alpha\beta\mu\nu}$, (ii)
$v^\mu$, $n^\mu$ and (iii)
\begin{eqnarray}
\sqrt{2} \frac{m_T}{E} \{ \veps(\lambda)^{*\mu\nu} v_\nu -
[\veps(\lambda)^*_{\alpha\beta} v^\alpha v^\beta ] n^\mu \} &=& \begin{cases}
0              & \mbox{for } \lambda = \pm 2, \\
\veps(\pm)^\mu & \mbox{for } \lambda = \pm1, \\
0              & \mbox{for } \lambda = 0,
\end{cases}\label{1st-vec}
\\
\sqrt{2} \frac{m_T}{E} \eps^{\mnrs}
 [ \veps(\lambda)^*_{\nu\alpha} v^\alpha ] n_\rho v_\sigma
&=& \begin{cases}
0                               & \mbox{for }\lambda = \pm 2, \\
\eps^{\mu\nu\rho\sigma}
 \veps(\pm)_\nu n_\rho v_\sigma & \mbox{for } \lambda = \pm1, \\
0                               & \mbox{for } \lambda = 0,
\end{cases}\label{2nd-vec}
\\
\sqrt{\frac{3}{2}} \left(\frac{m_T}{E}\right)^2
 [\veps(\lambda)^*_{\alpha\beta}v^\alpha v^\beta] n^\mu &=& \begin{cases}
0     & \mbox{for } \lambda = \pm 2, \\
0     & \mbox{for } \lambda = \pm1, \\
n^\mu & \mbox{for } \lambda = 0,
\end{cases}\label{3rd-vec}
\\
\sqrt{\frac{3}{2}} \left(\frac{m_T}{E}\right)^2 [ \veps(\lambda)^*_{\alpha\beta}v^\alpha v^\beta ]
 v^\mu
&=& \begin{cases}
0     & \mbox{for } \lambda = \pm 2, \\
0     & \mbox{for } \lambda = \pm1, \\
v^\mu & \mbox{for } \lambda = 0,
\end{cases}\label{4th-vec}
\end{eqnarray}
to project the relevant polarization states of the tensor meson. Eqs.
\eqref{1st-vec}, \eqref{3rd-vec} and \eqref{4th-vec} are the vectors, but
Eq.~\eqref{2nd-vec} the axial-vector. Matching the parities of the matrix
elements and using the three classes of the Lorentz structures, we can then
easily parametrize the form factors in the following results:
\begin{eqnarray}
\bra{T}|V^\mu|\ket{\barB} &=&
 -i 2E \left(\frac{m_T}{E}\right) \zetav_\perp \veps^{*\mnrs}
  v_\nu n_\rho  \veps^*_{\sigma\beta}v^\beta,
\label{LEETFF-vector}
\\
\bra{T}|A^\mu|\ket{\barB} &=&  2E \left(\frac{m_T}{E}\right) \zetaa_\perp
\left[
\veps^{*\mu\alpha}v_\alpha - (\veps^*_{\alphabeta}v^\alpha v^\beta) n^\mu
\right]
\nonumber\\&&
+ 2E \left(\frac{m_T^2}{E^2}\right)
(\veps^*_{\alphabeta}v^\alpha v^\beta)
 \left[\zetaa_{\para} n^\mu + \zetaa_{\para,1} v^\mu \right],
\\
\bra{T}|T^{\mu\nu}|\ket{\barB} &=& 2E
 \left(\frac{m_T^2}{E^2}\right)
 \zetat_\para \eps^{\mnrs} (\veps^*_{\alphabeta}v^\alpha v^\beta)  v_\rho n_\sigma
\nonumber\\&&
+2E \left(\frac{m_T}{E}\right)
 \zetat_\perp \eps^{\mnrs} n_\rho [ \veps^*_{\sigma\alpha} v^\alpha
 - (\veps^*_{\alphabeta}v^\alpha v^\beta)n_\sigma ]
\nonumber\\&&
+2E \left(\frac{m_T}{E}\right)\zetat_{\perp,1}
 \eps^{\mnrs} v_\rho [\veps^*_{\sigma\alpha}v^\alpha
 - (\veps^*_{\alpha\beta}v^\alpha v^\beta)n_\sigma],
\label{LEETFF-tensor}
\\
\bra{T}|T_5^{\mu\nu}|\ket{\barB} &=&
 - i2E \left(\frac{m_T}{E}\right) \zetatf_{\perp,1}
 \left\{
 \left[\veps^{*\mu\alpha}v_\alpha - (\veps^*_{\alpha\beta}v^\alpha v^\beta)n^\mu \right]
 v^\nu - (\mu \leftrightarrow \nu)
 \right\}
\nonumber\\&&
-i2E \left(\frac{m_T}{E}\right) \zetatf_{\perp} \left\{
\left[ \veps^{*\mu\alpha} v_\alpha - (\veps^*_{\alphabeta}v^\alpha v^\beta) n^\mu \right]
 n^\nu - (\mu \leftrightarrow \nu)
\right\}
\nonumber\\&&
-i2E \left(\frac{m_T^2}{E^2}\right)
 \zetatf_{\para} (\veps^*_{\alphabeta}v^\alpha v^\beta) (n^\mu v^\nu - n^\nu v^\mu),
\label{LEETFF-axialtensor}
\end{eqnarray}
where $\eps^{0123} = -1$ is adopted. $\bra{T}|T^{\mu\nu}|\ket{\barB}$ is
related to $\bra{T}|T_5^{\mu\nu}|\ket{\barB}$ by using the relation: $
\sigma^{\mu\nu}\gamma_5 \eps_{\mnrs} = 2i \sigma_{\rho\sigma}$. Note that for
the tensor meson only the states with helicities $\pm1$ and $0$ contribute to
the $\barB \to T$ transition in the LEET limit. $\zeta_\perp$'s are relevant to
$T$ with helicity $=\pm1$, and $\zeta_\para$'s to $T$ with helicity $=0$.

In order to reduce the number of the independent $\barB\to T$ form factors, we
consider the effective current operator $\qbar_n \Gamma b_v$ (with $\Gamma =
1,\gamma_5,\gamma^\mu,\gamma^\mu \gamma_5,
\sigma^{\mu\nu},\sigma^{\mu\nu}\gamma_5$) in the LEET limit, instead of the
original one $\qbar \Gamma b$ \cite{Charles:1998dr}. Here $b_v$ and $q_n$
satisfy $\slashchar{v} b_v = b_v$, $\slashchar{n} q_n = 0$ and
$(\slashchar{n}\slashchar{v}/2)q_n = q_n$. Employing the Dirac identities
\begin{eqnarray}
\frac{\slashchar{v}\slashchar{n}}{2} \gamma^\mu
 &=& \frac{\slashchar{v}\slashchar{n}}{2} \left(n^\mu \slashchar{v}
  - i \eps^{\mnrs}v_\nu n_\rho \gamma_\sigma \gamma_5 \right),
\\
\frac{\slashchar{v}\slashchar{n}}{2} \sigma^{\mu\nu}
 &=& \frac{\slashchar{v}\slashchar{n}}{2}
 \left[i(n^\mu v^\nu - n^\nu v^\mu) -i(n^\mu \gamma^\nu - n^\nu \gamma^\mu)
 \slashchar{v} - \eps^{\mnrs}v_\nu n_\rho \gamma_\sigma \gamma_5 \right],
\end{eqnarray}
where $\eps^{0123}=-1$ is adopted, one can obtain the following relations:
\begin{eqnarray}
\qbar_n b_v &=& v_\mu \qbar_n \gamma^\mu b_v, \label{relation-scalar}
\\
\qbar_n \gamma^\mu b_v &=& n^\mu \qbar_n b_v -i \eps^{\mnrs}
 v_\nu n_\rho \qbar_n \gamma_\sigma \gamma_5 b_v,
\\
\qbar_n \gamma^\mu\gamma_5 b_v &=& - n^\mu \qbar_n \gamma_5 b_v
 - i \eps^{\mnrs}  v_\nu n_\rho \qbar_n \gamma_\sigma b_v,
\\
\qbar_n \sigma^{\mu\nu} b_v &=& i \left[
 n^\mu v^\nu \qbar_n b_v - n^\mu \qbar_n \gamma^\nu b_v
 - (\mu \leftrightarrow \nu) \right]
 - \eps^{\mnrs} v_\rho n_\sigma \qbar_n \gamma_5 b_v,
\\
\qbar_n \sigma^{\mu\nu}\gamma_5 b_v &=& i \left[
 n^\mu v^\nu \qbar_n \gamma_5 b_v + n^\mu \qbar_n \gamma^\nu \gamma_5 b_v
 - (\mu \leftrightarrow \nu) \right]
  - \eps^{\mnrs} v_\rho n_\sigma \qbar_n b_v.
 \label{relation-axialtensor}
\end{eqnarray}
Substituting the above results into
Eqs.~\eqref{LEETFF-vector}-\eqref{LEETFF-axialtensor}, we have
\begin{eqnarray}
\zetav_\perp = \zetaa_\perp = \zetat_\perp = \zetatf_\perp & \equiv & \zeta_\perp,
\\
\zetaa_{\para} = \zetat_\para = \zetatf_\para & \equiv & \zeta_\para,
\\
\zetaa_{\para,1} = \zetatf_{\perp,1} = \zetat_{\perp,1} &=& 0,
\end{eqnarray}
and thus find that there are only two independent components,
$\zeta_\perp(q^2)$ and $\zeta_\para(q^2)$, for the $B \to T$ transition in the
LEET limit. In the full theory, the $\barB(p_B) \to
\barktwo(p_{\ktwo},\lambda)$ form factors  are defined as follows,
\newcommand{\ep}{e}
\begin{eqnarray}
\bra{\barktwo(p_{\ktwo},\lambda)}|\sbar \gamma^\mu b|\ket{\barB(p_B)}
&=& - i \frac{2}{m_B + \mktwo} \tilde V^{\ktwo}(q^2)
\eps^{\mnrs} p_{B\nu} p_{\ktwo\rho} \ep^*_\sigma,
\\
\bra{\barktwo(p_{\ktwo},\lambda)}|\sbar \gamma^\mu \gamma_5 b|\ket{\barB(p_B)}
&=& 2 \mktwo \tilde A_0^{\ktwo}(q^2) \frac{\ep^* \cdot p_B}{q^2} q^\mu
\nonumber\\&&
+ (m_B + \mktwo) \tilde A_1^{\ktwo}(q^2)
\left[ \ep^{*\mu} - \frac{\ep^* \cdot p_B}{q^2} q^\mu \right]
\nonumber\\&&
- \tilde A_2^{\ktwo}(q^2) \frac{\ep^* \cdot p_B}{m_B + \mktwo}
\left[ p_B^\mu + p_{\ktwo}^\mu - \frac{m_B^2 - \mktwo^2}{q^2} q^\mu
\right],
\nonumber\\&&
\\
\bra{\barktwo(p_{\ktwo},\lambda)}|\sbar \sigma^{\mu\nu}q_\nu b|\ket{\barB(p_B)}
&=&
 2 \tilde T_1^{\ktwo} (q^2) \eps^{\mnrs} p_{B\nu} p_{\ktwo\rho} \ep^*_\sigma,
\\
\bra{\barktwo(p_{\ktwo},\lambda)}|\sbar \sigma^{\mu\nu}\gamma_5 q_\nu b|\ket{\barB(p_B)}
&=& -i \tilde{T_2}^{\ktwo} (q^2) \left[
(m_B^2 - \mktwo^2) \ep^{*\mu} - (\ep^* \cdot p_B)(p_B^\mu + p_{\ktwo}^\mu)
\right]
\nonumber\\&&
- i \tilde{T_3}^{\ktwo}(q^2) (\ep^* \cdot p_B)
\left[ q^\mu - \frac{q^2}{m_B^2 - \mktwo^2}(p_B^\mu + p_{\ktwo}^\mu)
\right],
\nonumber\\&&
\end{eqnarray}
where $\ep^\mu \equiv \veps^{\mu\nu}(p_{\ktwo},\lambda) p_{B,\nu}/m_B$
corresponding to $\lambda=0,\pm1$. We have $ \ep^\mu  = (\absp / \mktwo)
\tilde\veps^\mu $, where $\tilde\veps(0) = \sqrt{2/3} \veps(0)$ and
$\tilde\veps(\pm1) = \sqrt{1/2} \veps(\pm1)$. We thus normalize these form
factors and obtain relations as follows
\begin{eqnarray}
\tilde A_0^{\ktwo}(q^2) \frac{\absp}{\mktwo} &\equiv&
 A_0^{\ktwo}(q^2)
\simeq \left(1-\frac{\mktwo^2}{m_B E}\right) \zeta_{\para}(q^2)
+ \frac{\mktwo}{m_B} \zeta_{\perp}(q^2),
\\
\tilde A_1^{\ktwo}(q^2) \frac{\absp}{\mktwo} &\equiv&
 A_1^{\ktwo}(q^2)
\simeq \frac{2E}{m_B + \mktwo} \zeta_{\perp}(q^2),
\\
\tilde A_2^{\ktwo}(q^2) \frac{\absp}{m_{\ktwo}} &\equiv&
 A_2^{\ktwo}(q^2)
\simeq \left(1+\frac{\mktwo}{m_B}\right) \left[
\zeta_{\perp}(q^2) - \frac{\mktwo}{E} \zeta_\para (q^2)
\right],
\\
\tilde V^{\ktwo}(q^2) \frac{\absp}{\mktwo} &\equiv&
 V^{\ktwo}(q^2)
\simeq \left(1+\frac{\mktwo}{m_B}\right) \zeta_{\perp}(q^2),
\\
\tilde T_1^{\ktwo}(q^2) \frac{\absp}{\mktwo} &\equiv&
 T_1^{\ktwo}(q^2)
\simeq \zeta_{\perp}(q^2),
\\
\tilde T_2^{\ktwo}(q^2) \frac{\absp}{\mktwo} &\equiv&
 T_2^{\ktwo}(q^2)
\simeq \left(1-\frac{q^2}{m_B^2 - \mktwo^2}\right) \zeta_{\perp}(q^2),
\\
\tilde T_3^{\ktwo}(q^2) \frac{\absp}{\mktwo} &\equiv&
 T_3^{\ktwo}(q^2)
\simeq \zeta_{\perp}(q^2) - \left(1-\frac{\mktwo^2}{m_B^2}\right)
\frac{\mktwo}{E} \zeta_\para(q^2),
\end{eqnarray}
where have used $\absp/E \simeq 1$.
Our results are consistent with Ref.~\cite{Datta:2007yk}.
Defining
\begin{eqnarray}
\tveps(0)^\mu 
=  \alphal \veps(0)^\mu,
\quad
\tveps(\pm1)^\mu 
=  \betat \veps(\pm1)^\mu,
\label{helicity-dependency}
\end{eqnarray}
we can easily generalize the studies of $B \to \kstar\gamma$, $B \to \kstar
\lpm$ and $B \to \kstar \nu \nubar$ to $B \to \ktwo \gamma$, $B \to \ktwo \lpm$
and $B \to \ktwo \nu \nubar$ processes. For the $\kstar$ cases, we have
$\alphal = \betat = 1$, whereas for the $\ktwo$ cases, we instead use $\alphal
= \sqrt{2/3}$ and $\betat = 1/\sqrt{2}$.

\section{Numerical Study}\label{NumericalStudy}

In the following numerical study, we use the input parameters listed in
\tbl\ref{tbl:inputs}. The Wilson coefficients that we adopt are the same as
that in Ref.~\cite{Hatanaka:2008gu}
\begin{table}[tbp]
\caption{Input parameters}\label{tbl:inputs}
\begin{ruledtabular}
\begin{tabular}{ll}
Tensor meson mass &
$m_{K_2^{*+}(1430)} = 1.426 \, \GeV$,
\quad
$m_{K_2^{*0}(1430)} = 1.432 \, \GeV$,
\\
$b$ quark mass \cite{Amsler:2008zzb}
&   $m_{b,{\rm pole}} = 4.79^{+0.19}_{-0.08} \, \GeV$,
\\
$B$ lifetime (picosecond)&
$\tau_{B^+} = 1.638$,
\quad
$\tau_{B^0} = 1.530$,
\\
CKM parameter \cite{ckmfitter}
& $|V_{ts}^* V_{tb}^{}| = 0.040\pm0.001$,
\quad
$|V_{ub}| = (3.44^{+0.22}_{-0.17}) \times 10^{-3}$.
\end{tabular}
\end{ruledtabular}
\end{table}

\subsection{$B\to\ktwo\gamma$ and $B\to\ktwo\lpm$}

The effective Hamiltonian relevant to the $B\to\ktwo\gamma$ and $B\to\ktwo\lpm$
decays is given by
\begin{eqnarray}
\calH_{\eff} &=& -
\frac{G_F}{\sqrt{2}} V_{tb} V_{ts}^* \sum_{i=1}^{10} c_i(\mu) \calO_i(\mu) + \hc,
\\
\calO_7 &=& - \frac{g_{em} m_b}{8\pi^2} \sbar \sigma_{\mu\nu}(1+\gamma_5)b F_{\mu\nu},
\quad
\calO_8 = - \frac{g_s m_b}{8\pi^2} \sbar_i \sigma_{\mu\nu}(1+\gamma_5) b_j G^{\mu\nu} T^{ij},
\nonumber \\
\calO_{9} &=& \frac{\alphaem}{2\pi} \sbar(1-\gamma_5)b (\lbar\l),
\quad
\calO_{10} = \frac{\alphaem}{2\pi} \sbar(1-\gamma_5)b (\lbar\gamma_5\l).
\end{eqnarray}
In analogy to $B\to\kstar\gamma$
\cite{Ali:1999mm,Ali:2001ez,Beneke:2001at,Bosch:2001gv}, the $B\to \ktwo
\gamma$ decay width reads
\begin{eqnarray}
\lefteqn{\Gamma(B\to\ktwo\gamma)}&& \nonumber\\
&=&
\frac{G_F^2 \alphaem\left|V_{ts}^* V_{tb}^{}\right|^2}{32\pi^4}
m_{b,{\rm pole}}^2 \mB^3 \left(1- \frac{\mktwo^2}{\mB^2}\right)^3
\left| c_7^{(0){\eff}} + A^{(1)} \right|^2
\left|T_1^{\ktwo}(0)\right|^2 \betat^2,
\end{eqnarray}
with $\betat = \sqrt{1/2}$.
Here $A^{(1)}$ is decomposed into the following components \cite{Ali:2001ez}
\begin{eqnarray}
A^{(1)}(\mu) &=& A_{c_7}^{(1)}(\mu) + A_{\rm ver}^{(1)}(\mu)
= -0.038 - 0.016 i .
\end{eqnarray}
\par
In the LEET limit, $T_1^{\ktwo}(q^2)$ can be parametrized in terms of two
independent functions $\zeta_\perp(q^2)$ and $\zeta_\para(q^2)$. Using
$c_7^{(0)\eff} = -0.315$ and the $\Br(B^0 \to \ktwoz\gamma)$ data in
\tbl\ref{ktwo-exp}, we estimate the value of $\zeta_\perp(0)$ as
\begin{eqnarray}
T_1^{\ktwo}(0) \simeq \zeta_\perp(0) &=& 0.27 \pm 0.03 {}^{+0.00}_{-0.01},
\label{zetaTvalue}
\end{eqnarray}
where the errors are due to the uncertainties of the experimental data and pole
mass of the $b$-quark, respectively. The uncertainty is mainly due to the error
of the data. We use the QCD counting rules to analyze the $q^2$-dependence of
form factors \cite{Chernyak:1983ej}. We consider the Breit frame, where the
initial $B$ meson moves in the opposite direction but with the same magnitude
of the momentum compared with the final state $K_2^*$, i.e., $\vec{p}_B =
-\vec{p}_{K_2^*}$. In the large recoil region, where $q^2 \sim 0$, since the
two quarks in mesons have to interact strongly with each other to turn around
the spectator quark, the transition amplitude is dominated by the
one-gluon exchange between the quark pair and is therefore proportional to $1/E^2$. Thus we get $\langle
K_2^*(p_{K_2^*},\pm 1)|V^\mu|B(p_B)\rangle \propto \eps^{\mu\nu\rho\sigma}
p_{B\nu} p_{K_2^* \rho} \veps(\pm )_\sigma \times 1/E^2 $ and  $\langle
K_2^*(p_{K_2^*},0)|A^\mu|B(p_B)\rangle \propto p_{K_2^*}^\mu \times 1/E^2$. In
other words, we have $\zeta_{\perp,\parallel}(q^2)\sim 1/E^2$ in the large
recoil region. Motivated by the above analysis, we will
model the $q^2$ dependence of the form-factor functions to be $
\zeta_{\perp,\parallel}(q^2) =
\zeta_{\perp,\parallel}(0)\cdot(1-q^2/m_B^2)^{-2} $. For the value\footnote{The light-front results infer that $\zeta_\perp$ and $\zeta_\parallel$ are of the same sign \cite{HC}.} of $\zeta_\parallel(0)$, within the framework of the SM model, it was shown that $f_T/f_L \approx 3(m_\phi/m_B)^2 (\zeta_\perp/\zeta_\parallel)^2$ for the $B\to \phi K_2^*$ decay \cite{Datta:2007yk}, where $f_T$ and $f_L$ are the
transverse and longitudinal components, respectively\footnote{Here the new-physics contribution can be negligible if it mainly affects $c^{\rm eff}_7$.}. Comparing with the
current data $f_L=0.80\pm 0.10$ for $B^+\to \phi K_2^*(1430)^+$ and
$f_L=0.901^{+0.059}_{-0.061}$ for $B^0\to \phi K_2^*(1430)^0$ \cite{HFAG}, we therefore parametrize
\begin{eqnarray}
\xi \equiv \zeta_\para(0)/\zeta_\perp(0),
\quad
\mbox{with } 0.8 \le \xi \le 1.2,
\label{xiValue}
\end{eqnarray}
to take into account the possible uncertainty.
%
\newcommand{\hats}{\hat{s}}
\newcommand{\us}{u(s)}
\newcommand{\hatu}{\hat{u}}
\newcommand{\hatus}{\hatu(\hats)}
\newcommand{\hatmktwo}{\hat{m}_{\ktwo}}
\newcommand{\hatml}{\hat{m}_{\l}}
\newcommand{\hatmb}{\hat{m}_b}

The invariant amplitude of $\barB \to \barktwo \l^+\l^-$, in analogy to
\cite{Ali:1999mm}, is given by
\begin{eqnarray}
\calM &=&
-i \frac{G_F \alphaem}{2\sqrt{2} \pi} V_{ts}^* V_{tb}^{} m_B
\left[ \calT_\mu \sbar \gamma^\mu b
     + \calU_\mu \sbar \gamma^\mu\gamma_5 b
\right],
\end{eqnarray}
where
\begin{eqnarray}
\calT_\mu &=&
 \calA \eps_{\mnrs} \tveps^{*\nu} p_B^\rho p_T^\sigma
 -i m_B^2 \calB \tveps^*_\mu
 +i \calC (\tveps^* \cdot p_B) p_\mu
 +i \calD (\tveps^* \cdot p_B) q_\mu ,
\\
\calU_\mu &=&
 \calE \eps_{\mnrs} \tveps^{*\nu} p_B^\rho p_T^\sigma
 -i m_B^2 \calF \tveps^*_\mu
 +i \calG (\tveps^* \cdot p_B) p_\mu
 +i \calH (\tveps^* \cdot p_B) q_\mu .
\end{eqnarray}
The $\calD$-term vanishes when equations of motion of leptons are taken
into account.
The building blocks $\calA,\cdots,\calH$ are given by
\begin{eqnarray}
\calA &=& \frac{2}{1+\hatmktwo} c_9^{\eff} V^{\ktwo}(s) +
\frac{4\hatmb}{\hats} c_7^{\eff} T_1^{\ktwo}(s),
\\
\calB &=& (1+\hatmktwo) \left[
 c_9^{\eff}(\hats)  A_1^{\ktwo}(s) + 2 \frac{\hatmb}{\hats} (1-\hatmktwo)
 c_7^{\eff} T_2^{\ktwo}(s)
\right],
\\
\calC &=& \frac{1}{1-\hatmktwo} \biggl[
 (1-\hatmktwo) c_9^\eff(\hats) A_2^{\ktwo}(s) + 2\hatmb c_7^\eff
 \left(T_3^{\ktwo}(s) + \frac{1-\hatmktwo}{\hats}T_2^{\ktwo}(s)\right)
\biggr],
\\
\calD &=& \frac{1}{\hats}
\biggl[
 c_9^\eff(\hats) \{(1+\hatmktwo) A_1^{\ktwo}(s) - (1-\hatmktwo)A_2^{\ktwo}(s)\}
\nonumber\\&&
\phantom{MMM} - 2\hatmktwo A_0^{\ktwo}(s)
 - 2\hatmb c_7^\eff T_3^{\ktwo}(s)
\biggr],\\
\calE&=& \frac{2}{1+\hatmktwo} c_{10} V^{\ktwo}(s),
\quad
\calF = (1 + \hatmktwo) c_{10} A_1^{\ktwo}(s),
\quad
\calG = \frac{1}{1+\hatmktwo} c_{10} A_2^{\ktwo}(s),
\\
\calH &=& \frac{1}{\hats} c_{10}\left[
 (1+\hatmktwo)A_1^{\ktwo}(s) - (1-\hatmktwo) A_2^{\ktwo}(s) - 2\hatmktwo A_0^{\ktwo}(s)
\right],
\end{eqnarray}
where $\hat s \equiv s / m_B^2$ and $s \equiv (p_+ + p_-)^2$ with $p_\pm$ being
the momenta of the leptons $\l^\pm$. $c_9^\eff(\hats) = c_9 + Y_{\rm
pert}(\hats) + Y_{\rm LD}$ contains both the perturbative part $Y_{\rm
pert}(\hats)$ and long-distance part $Y_{\rm LD}(\hats)$.  $Y(\hats)_{\rm LD}$
involves $B \to K_2^* V(\cbar c)$ resonances, where $V(\cbar c)$ are the vector
charmonium states. We follow Refs.~\cite{Ali:1991is,Lim:1988yu} and set
\begin{eqnarray}
Y_{\rm LD}(\hats)
&=&
 - \frac{3\pi}{\alphaem^2} c_0
\sum_{V = \psi(1s),\cdots} \kappa_V \frac{\hat m_V \Br(V\to
\l^+\l^-)\hat{\Gamma}_{\rm tot}^V}{\hat s - \hat m_V^2 + i \hat m_V
\hat{\Gamma}_{\rm tot}^V},
\end{eqnarray}
where $\hat{\Gamma}_{\rm tot}^V \equiv \Gamma_{\rm tot}^V/\mB$ and $\kappa_V
=2.3$. The detailed parameters used in this paper can be found in
Ref.~\cite{Hatanaka:2008gu}.
The longitudinal, transverse and total differential decay rates are
respectively given by
\begin{eqnarray}
\frac{d\Gamma_L}{d s}
\equiv\left. \frac{d\Gamma}{d s} \right|_{\substack{\alphal=\sqrt{2/3}\\ \betat=0}},
\quad
\frac{d\Gamma_T}{d s}
\equiv\left. \frac{d\Gamma}{d s} \right|_{\substack{\alphal=0\\ \betat=\sqrt{1/2}}},
\quad
\frac{d\Gamma_{\rm total}}{d s}
\equiv\left. \frac{d\Gamma}{d s} \right|_{\substack{\alphal=\sqrt{2/3}\\ \betat=\sqrt{1/2}}}.
\end{eqnarray}
with
\begin{eqnarray}
\frac{d\Gamma}{d\hats}
&=&
\frac{G_F^2 \alphaem^2 m_B^5}{2^{10}\pi^5}|V_{ts}^*V_{tb}|^2
\Biggl\{
\nonumber\\&&
\frac{1}{6}|\calA|^2 \hatus \hats \betat^2
\left\{
 3 \left[ 1- 2 (\hatmktwo^2 + \hats) + (\hatmktwo^2 - \hats)^2
 \right] - \hatus^2
\right\}
+\betat^2  |\calE|^2 \hats \frac{\hatus^3}{3}
\nonumber\\&&
+ \frac{1}{12 \hatmktwo^2 \lambda} |\calB|^2 \hatus
\left\{ 3
 \left[ 1 - 2(\hatmktwo^2 + \hats) + (\hatmktwo^2 - s)^2 \right]
 - \hatus^2 \right\}
\nonumber\\&& \phantom{MMMM} \times
\left[ (-1 + \hatmktwo^2 + \hats)^2 \alphal^2 + 8 \hatmktwo^2 \hats \betat^2
\right]
\nonumber\\&&
+ \frac{1}{12 \mktwo^2 \lambda} |\calF|^2 \hatus
\Bigl\{
3\alphal^2 \lambda^2 \nonumber\\&& \phantom{MMMM}
+ \hatus^2
 \bigl[
16 \hatmktwo^2 \hats \betat^2 - (1 - 2(\hatmktwo^2+\hats) + \hatmktwo^4 +
\hats^2 - 10\hatmktwo^2 \hats) \alphal^2 \bigr] \Bigr\} \nonumber\\&&
+ \alphal^2 \hatu(s) \frac{\lambda}{4 \hatmktwo^2}
\biggl[ |\calC|^2 \left(\lambda - \frac{\hatus^2}{3}\right)
+ |\calG|^2 \left(\lambda - \frac{\hatus^2}{3}
 + 4 \hatml^2 (2 + 2 \hatmktwo^2 - \hats)\right)\biggr]
\nonumber\\&&
- \alphal^2 \hatu(s) \frac{1}{2 \hatmktwo^2}   \biggl[
 \Re(\calB \calC^*) \left(\lambda - \frac{\hatus^2}{3}\right)(1 - \hatmktwo^2 - \hats)
\nonumber\\&&
\phantom{MMMM} +
 \Re(\calF \calG^*)\left\{
 \left(\lambda - \frac{\hatus^2}{3}\right)(1 - \hatmktwo^2 - \hats)
 + 4 \hatml^2 \lambda \right\}
\biggr]
\nonumber\\&&
-2 \alphal^2 \hatu(s) \frac{\hatml^2}{\hatmktwo^2} \lambda \left[
\Re(\calF \calH^*)  - \Re(\calG \calH^*)(1 - \hatmktwo^2)\right]
+ \alphal^2 \hatu(s) \frac{\hatml^2}{\hatmktwo^2} \hats \lambda  |\calH|^2
\Biggr\}.
\end{eqnarray}
We have chosen the kinematic variables $\hat u \equiv u/ m_B^2$ and $\hatu s
\equiv u(s) / m_B^2$, where $u = - u(s) \cos\theta$ and
\begin{eqnarray}
u(s) &\equiv& \sqrt{\lambda \left(1 - \frac{4\hatml^2}{\hats} \right)},
\end{eqnarray}
with
\begin{eqnarray}
\lambda &\equiv& 1 + \hatmktwo^4 + \hats^2
 - 2\hatmktwo^2 - 2\hats - 2\hatmktwo^2 \hats,
\end{eqnarray}
and $\theta$ being the angle between the moving direction of $\l^+$ and $B$ meson
in the center of mass frame of the $\l^+ \l^-$ pair.
\begin{figure}[tbp]
\caption{The differential decay rates $d\Gamma_{\rm total}
(B^0\to\ktwozv\mupm)/ds$ as functions of the dimuon invariant mass $s$. The
solid (dashed) curve corresponds to the center value of the decay rate with
(without) the charmonium resonance effects. }\label{fig:specplot}
\includegraphics{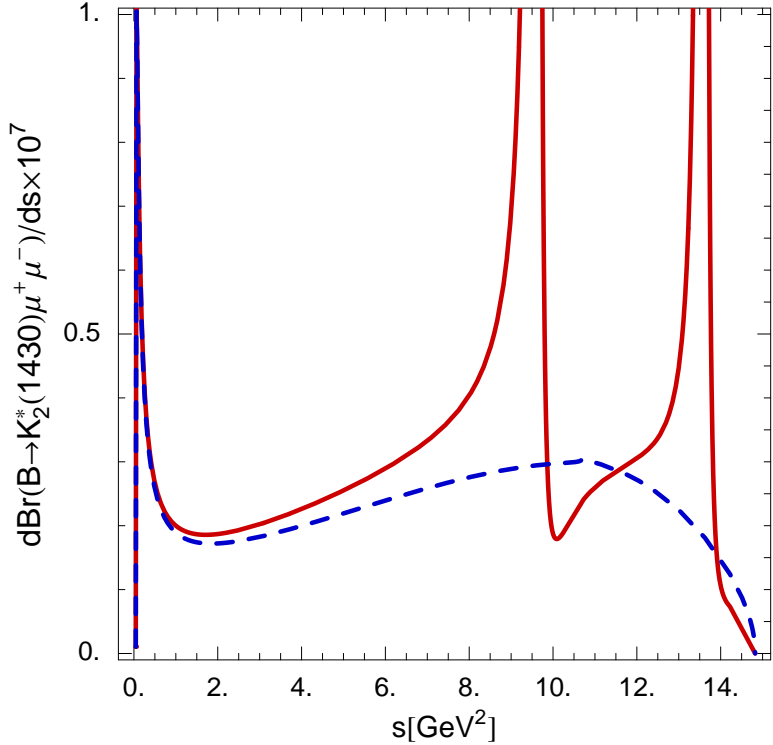}
\end{figure}
In \fig\ref{fig:specplot}, the total decay rates for $B \to \ktwov \mupm$with
and without charmonium resonances are plotted. The detailed results for the
charmonium resonances can be found in Refs.~\cite{Lim:1988yu,Ali:1991is}. The
branching fraction for nonresonant $B\to \ktwo\mu^+ \mu^-$ is obtained to be
\begin{eqnarray}
\Br(B^0\to\ktwozv\mu^+ \mu^-) &=&
(3.5 {}^{+1.1}_{-1.0} {}^{+0.7}_{-0.6}) \times 10^{-7},
\end{eqnarray}
where the first error comes from the variation of $\zeta_\perp$ in Eq.
\eqref{zetaTvalue}, the second error from the uncertainty of $\xi$ in
Eq.~\eqref{xiValue}.

\par
\begin{figure}[tbp]
\caption{Longitudinal fraction distributions $d F_L/ds$ as functions of $s$.
The thick (blue) and thin (red) curves correspond to the central values of
$B\to\ktwozv\mupm$ and $B^0\to K^{*0}(892)\mupm$ decays, respectively. The
solid and dashed curves correspond to the SM and new-physics model with flipped
sign of $c_7^{\eff}$, respectively. }\label{fig:FL}
\includegraphics{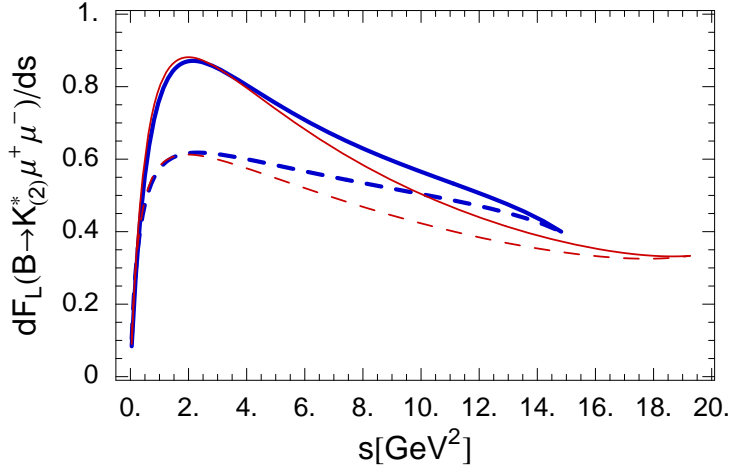}
\end{figure}

 
The longitudinal fraction distribution for $B\to\ktwo\l^+\l^-$ decay is defined
as
\begin{eqnarray}
\frac{d F_L}{ds} \equiv
\frac{d\Gamma_L}{ds} \Bigg/ {\frac{d\Gamma_{\rm total}}{ds}}.
\end{eqnarray}
In \fig\ref{fig:FL}, the longitudinal fraction distribution for the $B\to
\ktwov\mupm$ decay is plotted. For comparison, we also plot $F_L(B\to
K^*(892)\mupm)/ds$ as a benchmark. For small $s$ ($\lesssim 3\,\GeV^2$),
$B\to\kstar\mupm$ and $B\to\ktwo\mupm$ have similar rates for the longitudinal
fraction, while for large $s$ ($\gtrsim 4\,\GeV^2$) the $dF_L/ds$ for the
$B\to\ktwo\mupm$ decay slightly exceeds the $B\to\kstar\mupm$. More
interestingly, when $s \sim 3\,\GeV^2$, the result of the new-physics models
with the flipped sign solution for $c_7^\eff$ can deviate more remarkably from
the SM prediction (and can be reduced by $20-30$\%).

\par
\begin{figure}[tbp]
\caption{Forward-backward asymmetries $d\overline{A}_{\rm FB}/ds$ for $B\to
\ktwov \mupm$(thick curves) and $B \to K^*(892) \mupm$ (thin curves) as
functions of the dimuon invariant mass $s$. The solid and dashed curves
correspond to the SM and new-physics model with flipped sign of $c_7^\eff$.
Variation due to the uncertainty from $\zeta_{\para}(q^2)/\zeta_{\perp}(q^2)$
(see Eq.~\eqref{xiValue}) is denoted by dotted curves. }\label{fig:afb}
\includegraphics{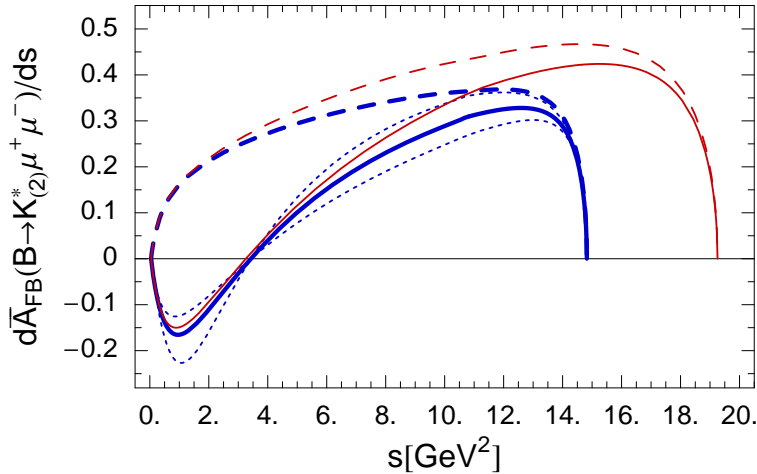}
\end{figure}

The forward-backward asymmetry for the $\barB \to \barktwo \l^+ \l^-$ decay is
given by
\begin{eqnarray}
\frac{d A_{\rm FB}}{d\hat{s}} &=&
 -  \betat^2 \frac{G_F^2\alphaem^2 m_B^5}{2^{10}\pi^5}
|V_{ts}^* V_{tb}|^2
\hat{s} \hat{u}(s)^2
\left[\Re(\calB \calE^*) + \Re(\calA \calF^*)\right]
\biggr|_{\betat = \sqrt{1/2}}
\\
&=&
 -  \betat^2 \frac{G_F^2\alphaem^2 m_B^5}{2^{10}\pi^5}
|V_{ts}^* V_{tb}|^2
\hat{s} \hat{u}(s)^2
\biggl[
\Re(c_{10}^{} c_9^{\eff}) V^{\ktwo} A_1^{\ktwo}
\nonumber\\&&\phantom{M}
+ \frac{\hatmb}{\hats} \Re(c_{10}^{} c_7^{\eff})
 \left\{
  (1-\hatmktwo) V^{\ktwo} T_2^{\ktwo} + (1+ \hatmktwo) A_1^{\ktwo} T_1^{\ktwo}
 \right\}
\biggr]\biggr|_{\betat=\sqrt{1/2}}.
\label{eq-afb}
\nonumber\\
\end{eqnarray}
In \fig\ref{fig:afb} we illustrate the normalized forward-backward asymmetry $d
\overline{A}_{\rm FB} / d s \equiv (d A_{\rm FB} / d s) / (d \Gamma_{\rm total}
/ d s)$ for $B\to\ktwo\mupm$ together with $B\to\kstar\mupm$.

In the SM, the forward-backward asymmetry zero $s_0$ for $B \to \ktwo \mupm$ is defined by
\begin{eqnarray}
\lefteqn{\Re [c_{10}^{} c_9^{\eff}(\hats_0) ] V^{\ktwo}(s_0) A_1^{\ktwo}(s_0) }
\nonumber&&\\
&=& - \frac{\hatmb}{\hats_0} \Re(c_{10}^{} c_7^{\eff})
 \left\{
  (1-\hatmktwo) V^{\ktwo}(s_0) T_2^{\ktwo}(s_0)
  + (1+ \hatmktwo) A_1^{\ktwo}(s_0) T_1^{\ktwo}(s_0)
 \right\}.
\end{eqnarray}
We obtain
\begin{eqnarray}
s_0 &=& 3.4 \pm 0.1\, \GeV^2,
\end{eqnarray}
where the error comes from the variation of $m_b$. This result is very close to
the zero for $B \to \kstar \mupm$. As shown in \fig\ref{fig:afb}, it is
interesting to note that the form factor uncertainty of the zero vanishes in
the LEET limit.

The asymmetry zero exists only for $\Re [c_9^\eff(s) c_{10}] \Re (c_7^\eff
c_{10}) < 0$. Therefore with the flipped sign of $c_7^\eff$ along, compared
with the SM prediction, the asymmetry zero disappears, and $d A_{\rm FB}/ds$ is
positive for all values of $s$. From recent measurements for $B\to\kstar \lpm$
decays, the solution with the flipped sign of $c_7^\eff$ seems to be favored by
the data \cite{Matias:2008uw,Eigen:2008nz,Bobeth:2008ij}. One can find the
further discussion in Ref.~\cite{Hatanaka:2008gu} for the $B \to K_1(1270)\lpm$
decays.

%
%

\subsection{$B\to \ktwo \nu \nubar$}

In the SM, $b \to s \nu \nubar$ proceeds through $Z$ penguin and box diagrams
involving top quark exchange \cite{Buras:1997fb}. One of the reasons that we
are interested in the study of decays going through $b \to s \nu \nubar$ is the
absence of long-distance corrections related to the relevant four-fermion
operators. Moreover, the branching fractions are enhanced by the summation over
three light neutrinos. New physics contributions arising from new loop and/or
box diagrams may significantly modify the predictions. In the SM, the branching
fractions involving $K$ or $\kstar$ are predicted to be $\Br(B \to K \nu\nubar)
\simeq 3.8 \times 10^{-6}$ and $\Br(B \to \kstar \nu\nubar) \simeq 13 \times
10^{-6}$ \cite{Buras:1997fb,Ali:1991fy}, while only upper limits $10^{-4} \sim
10^{-5}$ were set in the experiments \cite{:2007zk,:2008fr,Eigen:2008nz}. In
the new physics scenario, the contribution originating from the nonstandard
$Z^0$ coupling can enhance the branching fraction by a factor ten
\cite{Buchalla:2000sk}. This mode is also relevant to search for light dark
matter \cite{Bird:2004ts} and unparticles \cite{Georgi:2007ek,Aliev:2007gr}.

\begin{figure}[tbp]
\caption{Branching fraction distribution $d\Br(B^0 \to \ktwoz\nubar\nu)/ds$ as
a function of the missing invariant mass squared $s$ within the SM.
The solid (black), dashed (blue), dotted (green) and dot-dashed (red) curves
correspond to the total decay rate and the polarization rates with helicities
$h=0$, $-1$, $+1$, respectively.}\label{missing-plot}
\includegraphics{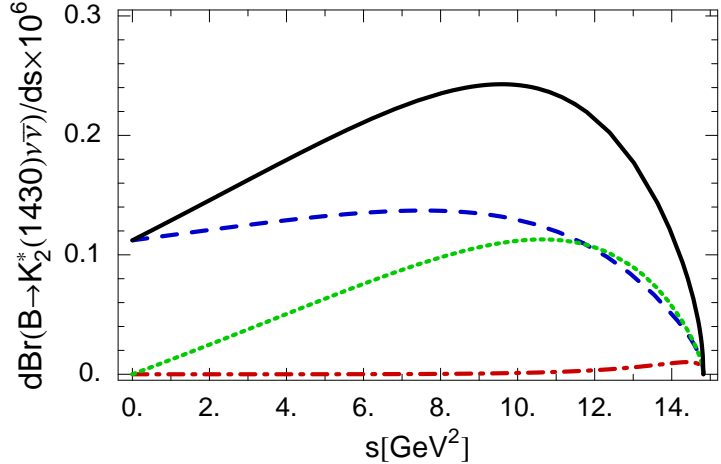}
\end{figure}

The generally effective weak Hamiltonian relevant to the $b \to s \nu\nubar$ decay is given by
\begin{eqnarray}
\calH_{\eff} =
c_L \sbar \gamma^\mu(1-\gamma_5) b\, \nubar \gamma_\mu (1-\gamma_5) \nu +
c_R \sbar \gamma^\mu(1+\gamma_5) b\, \nubar \gamma_\mu (1-\gamma_5) \nu,
\end{eqnarray}
where $c_L$ and $c_R$ are left- and right-handed weak hadronic current
contributions, respectively. New physics effects can modify the SM value of
$c_L$, while $c_R$ only receives the contribution from physics beyond the SM
\cite{Buchalla:2000sk}. In the SM we have
\begin{eqnarray}
c_L^{SM} &=&
\frac{G_F}{\sqrt{2}} \frac{\alpha_{EM}}{2\pi\sin^2\theta_W}
V_{tb}^{} V_{ts}^* X(x_t) = 2.9 \times 10^{-9},
\end{eqnarray}
where the detailed form of $X(x_t)$ can be found in Refs.~\cite{Inami:1980fz,Buchalla:1993bv,Buchalla:1995vs,Buchalla:1998ba}.
The $\ktwo$ helicity polarization rates of the missing invariant mass-squared distribution $d\Gamma_h/ds$ of the $B\to\ktwo\nubar\nu$ decay are given by \cite{Colangelo:1996ay,Melikhov:1998ug,Kim:1999waa,Buchalla:2000sk},
\begin{eqnarray}
\frac{d\Gamma_0}{d \hats} &=& 3\alphal^2
\frac{|\vec{p}|}{48\pi^3}  \frac{|c_L-c_R|^2}{\mktwo^2}
\nonumber\\&& \phantom{iii}
\times
\left[
(m_B + \mktwo)(m_B E - \mktwo^2) A_1^{\ktwo}(q^2)-
\frac{2m_B^2}{m_B+\mktwo}|\vec{p}|^2 A_2^{\ktwo}(q^2)
\right]^2,
\\
\frac{d\Gamma_{\pm1}}{d \hats} &=& 3\betat^2
\frac{|\vec{p}| q^2}{48\pi^3}
\nonumber\\&&
\times
 \biggl|
 (c_L + c_R) \frac{2m_B |\vec{p}|}{m_B + \mktwo} V^{\ktwo}(q^2) \mp
 (c_L - c_R) (m_B + \mktwo) A_1^{\ktwo}(q^2)
 \biggr|^2
,
\end{eqnarray}
where $\hats \equiv s/m_B^2$, $\alphal=\sqrt{2/3}$ and $\betat=\sqrt{1/2}$ with
$0 \le s \le (m_B - \mktwo)^2$ being the invariant mass squared of the
neutrino-antineutrino pair. Here the factor $3$ counts the numbers of the
neutrino generations. $\vec{p}$ and $E$ are the three-momentum and energy of
the $\ktwo$ in the $B$ rest frame. In \fig\ref{missing-plot}, we show the
distribution of the missing invariant mass-squared for the
$B\to\ktwov\nubar\nu$ decay within the SM.
We find
\begin{eqnarray}
\Br(B^0\to\ktwozv\nubar\nu) = (2.8^{+0.9}_{-0.8} {}^{+0.6}_{-0.5}) \times 10^{-6},
\end{eqnarray}
where the first and second errors are due to the uncertainty of the form factors and $\xi$, respectively.
%
\section{Summary}\label{Summary}

We have studied the radiative and semileptonic $B$ decays involving the tensor
meson $\ktwov$ in the final states. Using the large energy effective theory
techniques, $B \to \ktwov$ transition form factors have been formulated in the
large recoil region. There are only two independent functions
$\zeta_\perp(q^2)$ and $\zeta_\para(q^2)$ that describe all relevant form
factors. We have determined the value of $\zeta_\perp(0)$ from the measurement
of $\Br(B^0\to\ktwozv\gamma)$.
Adopting a dipole $q^2$-dependency for the LEET functions and
$\zeta_\para(q^2)/\zeta_\perp(q^2) = 1.0 \pm 0.2$, for which the former
consists with the QCD counting rules and the latter is favored by the $B\to
\phi K_2^*$ data, we have investigated the decays $B \to \ktwo \lpm$ and $B \to
\ktwo \nu\nubar$. Note that $\zeta_\parallel$ only gives corrections of order
$m_{K_2^*}/m_B$. We have discussed two dedicated observables, the longitudinal
distribution $dF_L/ds$ and forward-backward asymmetry, in the $B \to K_2^*
\ell^+ \ell^-$ decay. Recent forward-backward asymmetry measurements for $B
\to\kstar\lpm$ decays \cite{Ishikawa:2006fh,:2008ju,Adachi:2008sk} seem to (i)
allow the possibility of flipping the sign of $c_7^{\eff}$, or (ii) have both
$c_9$ and $c_{10}$ flipped in sign, as compared with the SM. Meanwhile, in the
large recoil region, \BABAR{} has recently reported the large isospin asymmetry
for  the $B \to K^* \ell^+ \ell^-$ decays, which qualitatively favors the
flipped sign $c_7^{\eff}$ model over the SM \cite{Eigen:2008nz}. Therefore in
the present study, in addition to the SM, we focus the new-physics effects on
$c_7^{\eff}$ with the sign flipped. It should be note that the magnitude of
$c_7^{\eff}$ is stringently constrained by the $B\to X_s\gamma$ data which is
consistent with the SM prediction.

For the $B \to \ktwo \lpm$ decay, of particular interest is the large recoil
region, where the uncertainties of form factors are considerably reduced not
only by taking the ratios of the form factors but also by computing in the
large $E_{K_2^*}$ limit. In this region, where the invariant mass of the lepton
pair $s \simeq 2-4\,\GeV^2$, due to the flipped sign of  $c_7^{\eff}$ compared
with the SM result, $d F_L/d s$ is reduced by $20-30$\%, and its value can be
$\sim 0.8$. One the other hand, in the SM the asymmetry zero is about
$3.4\,\GeV^2$, but changing the sign of $c_7^{\eff}$ yields a positive
forward-backward asymmetry for all values of the invariant mass of the lepton
pair.

We have obtained the branching fraction for $B \to \ktwo \nu \nubar$ in the SM. This mode enhanced by the summation over three light neutrinos is theoretically cleaner due to the absence of long-distance corrections related to the relevant four-fermion operators. This decay is relevant for the search for the nonstandard $Z^0$ coupling, light dark matter and unparticles.

In summary, the investigation of the semileptonic $B$ decays involving $\ktwov$
will further provide complementary  information on physics beyond the standard
model. Our results also exhibit the impressed resemblance of  the physical
properties between $B\to K_2^*(1430) \lpm, \nu\nubar$ and $B\to K^*(892) \lpm,
\nu\nubar$.
%
\begin{acknowledgments}
This research was supported in parts by the National Science Council of R.O.C.
under Grant No.\ NSC96-2112-M-033-004-MY3 and No.\ NSC97-2811-033-003 and by
the National Center for Theoretical Science.
\end{acknowledgments}


\end{document}